\documentclass[useAMS]{mn2e}
\usepackage{times}
\input epsf.sty

\def\rtr{r_{\rm tr}}
\def\Rg{R_{\rm g}}
\def\RS{R_{\rm Sch}}

\def\NH{N_{\rm H}}
\def\lh{l_{\rm h}}

\def\lhz{l_{\rm h,0}}

\def\lsoft{l_{\rm s}}

\def\Or{\Omega/2\pi}

\def\taumin{\tau_{\rm min}}
\def\taumax{\tau_{\rm max}}
\def\tmax{t_{\rm max}}

\def\taues{\tau_{\rm es}}

\def\sigmaT{\sigma_{\rm T}}
\def\me{m_{\rm e}}
\def\kT{k T_{\rm e}}
\def\MSun{{\rm M}_{\odot}}

\def\Ka{K$\alpha$\ }

\def\Aqpo{A_{\rm qpo}}
\def\Tqpo{T_{\rm qpo}}
\def\Tsoft{T_{\rm soft}}
\def\fqpo{f_{\rm qpo}}
\def\phiqpo{\phi_{\rm qpo}}

\def\rstart{r_{\rm i}}
\def\rend{r_{\rm f}}
\def\fba{f_{b,1}}
\def\fbb{f_{b,2}}

\title[X--ray quasi-periodic oscillations]
{Modelling the energy dependencies of X-ray quasi-periodic oscillations in 
accreting compact objects}

\author[P. T. \.{Z}ycki and M. Sobolewska]{Piotr T. \.{Z}ycki\thanks{e-mail: ptz@camk.edu.pl} and Ma{\l}gorzata Sobolewska \\
    Nicolaus Copernicus Astronomical Center, Bartycka 18, 00-716 Warsaw, Poland}

\date{8 September 2005}

\begin{document}
\label{firstpage}

\maketitle

\begin{abstract}

We have constructed models of quasi-periodic variability of X--ray emission
from accreting compact objects. Assuming a general scenario of a propagation
model of variability, with inverse Compton upscatering as the emission mechanism,
we have considered a number of cases for the periodic modulation: modulation
of the plasma heating rate, cooling rate by external soft photons, and the 
amplitude of the reprocessed component. We have computed various 
observational characteristics which can be compared to good quality data.
These include Fourier-frequency resolved spectra and results of cross-correlation
analysis between light-curves at different energies. Each model of modulation 
predicts specific observational signatures, which help in identifying
the physical processes driving QPO emission in accreting sources.

\end{abstract}


\section{Introduction}

X--ray lightcurves from accreting compact objects (neutron star X--ray binaries,
black hole X--ray binaries and active galactic nuclei) are generally
non-periodic, showing significant variability in broad range of time scales.
Consequently, their power spectra are broad, extending through  a few decades
in Fourier frequency (van der Klis 1995; Markowitz et al.\ 2003). 
Nevertheless, periodic, or quasi-periodic components appear
often in the lightcurves of the stellar systems. Typically, these quasi-periodic
oscillations (QPO) contain only a small fraction of total variability power,
and are hardly, if at all, recognizable in lightcurves in the time domain. 
Usually, they are
only clearly visible in power density spectra (PDS), where a large number
of oscillations is co-added. The observed QPO frequencies  range from Keplerian
frequency near a surface of a neutron star (kilo-Hz QPO in some neutron star
X--ray binaries) to below $\sim 0.1$ Hz (review in Wijnands 2001). 

Despite the fact that a QPO component is almost never the dominant light curve
component, their significance for our understanding of the accretion process
should not be underestimated. 
In accreting black hole systems QPO features seem to appear preferentially when
the state of the source is changing. For example, Rutledge et al.\ (1999) found
correlations between spectral state transitions and QPO appearance in soft X--ray
transient sources in {\it Ginga\/} data. Numerous works were devoted to study
such correlations in good quality {\it RXTE\/} data (e.g., Cui et al.\ 1999;
Kalemci et al.\ 2004; Rodriguez et al.\ 2004).
In many neutron star X--ray binaries many long-lived QPO features were observed
and studies of their time history revealed such interesting phenomena as
saturation of QPO frequency with mass accretion rate (Kaaret et al.\ 1999)
or ``parallel tracks'' phenomenon (van der Klis 2001 and references therein). 
This latter phenomenon is a hysteresis effect similar to that observed also 
on longer time scales
in some black hole X-ray binaries (Maccarone \& Coppi 2003). QPO offer also
a method to constrain the mass of the central compact object, provided that
their frequency is identified with a physical frequency of the system.

A variety of  spectral/timing behaviour, including several types of QPO, 
is shown by the black hole binary (BHB) system GRS 1915+105.
This is a peculiar source, most likely due to its high mass accretion
rate (Sobolewska \& \.{Z}ycki 2004; Done, Wardzi\'{n}ski \& Gierli\'{n}ski 2004). 
Recently, Miller \& Homan (2005) noticed changes in the intensity of the Fe 
\Ka fluorescent line at 6.4 keV during QPO period in this source. This is a potentially
important clue to the origin of the periodicity, suggesting that
variations occur in the relative configuration of the cold reprocessing matter
and the X--ray source. 

All of these demonstrate the importance of understanding the origin of 
the QPO features. 
Of course, given the  variety of QPO components it is rather unlikely that
a single mechanism could explain all of them, despite correlations between
different QPO frequencies (Belloni, Psaltis \& van der Klis 2002).

QPO models considered so far concentrated on identifying the frequencies of
oscillations (see, e.g., Lee, Abramowicz \& Klu\'{z}niak 2004 and references 
therein, and a review of QPO models in Psaltis 2001). 
Rather little attention was paid to the fact that it is the X--ray emission,
which is modulated. A notable exception
is the paper by Giannios \& Spruit (2004), where a model of global oscillations
of the hot inner flow interacting with outer cool disc was constructed. 
Oscillations are driven by feedback loop between heating the cool disc by ions
from the hot flow and cooling of the hot disc by soft photons from the cool disc.
The model predicts that r.m.s. variability of the QPO is highest at lower energies,
therefore creating a rather soft QPO spectrum.

In this paper we address the problem of QPO origin from a complementary 
point of view, considering how such variable hard X--ray emission may be
generated. Assuming
that the X--ray are produced in the thermal (or hybrid thermal-nonthermal;
Coppi 1999) Comptonization process (review in Zdziarski \& Gierli\'{n}ski 2004),
we construct a number of geometrical/phenomenological scenarios of 
oscillations. The main idea here is that -- irrespectively of the physical
mechanism  -- QPO may be driven by modulations of
only a few physical parameters which determine the Comptonized spectrum:
plasma heating rate, cooling rate by soft photons,  
amplitude of the reflected component (feedback between the heating and 
cooling),
or a temperature of soft seed photons. It is also possible that a QPO is
a simple geometrical effect of, for example, periodic obscuration of the
source of emission. For each such case
we predict various energy and timing characteristics which could be
compared with observations. In particular, we emphasize the Fourier-frequency
resolved spectroscopy as an important tool (Miyamoto et al.\ 1991;
Revnivtsev, Gilfanov \& Churazov
1999; \.{Z}ycki 2002; 2003, hereafter Z03).

\section{The Model}
\label{sec:model}

The general scenario for time variability considered in this paper
is a radial propagation model, as described in details in Z03 and
\.{Z}ycki (2004).
That is, we assume that X--rays are produced
by structures propagating radially inwards, towards the central compact 
object (Kotov, Churazov \& Gilfanov 2001). 
Dissipation of gravitational energy causes a flare
of radiation as the structures are approaching the center. 
The X--rays are produced in inverse 
Compton process on thermal electrons, with the soft seed photons provided 
by optically thick cold accretion disc (review in Done 2002).
Evolution of energy spectrum during a flare
of radiation, from softer to harder, produces hard X-ray time lags --
commonly observed in these sources (Poutanen \& Fabian 1999; 
review in Poutanen 2001).
The model reproduces also the Fourier-frequency resolved spectra 
($f$-spectra; Revnivtsev et al.\ 1999), which get harder with increasing 
Fourier frequency, at least in low/hard state of BHB (Z03). 
We employ the idea of flare avalanches to generate
lightcurves with PDS of a doubly-broken power law
form (Stern \& Svensson 1996; Poutanen \& Fabian 1999). The correlations
between the flares (avalanches) go some way towards explaining the
flux-$\sigma$ relation of Uttley \& McHardy (2001).

We note that the propagation model was formulated for low/hard state of BHB,
whose spectral/timing properties are known best. QPO are mostly observed
in softer/brighter states of BHB. However, there is certainly a continuity
of spectral/timing properties when the sources make transitions between
the states (see, e.g., Zdziarski \& Gierli\'{n}ski 2004; Done \& Gierli\'{n}ski 2004;
Kubota \& Done 2004), so it does seem reasonable to adopt the same general 
scenario for the soft states as well. On the other hand, data analyses do show
differences between some timing properties in hard and soft states, for example
in Fourier-frequency resolved spectra (Revnivtsev et al.\ 1999 for hard
state; Sobolewska \& \.{Z}ycki 2005 for soft state).

The flare of radiation is described by a specific time dependence of plasma 
heating rate,
expressed as compactness parameter, $l\equiv (L/D) (\sigmaT/\me c^3)$,
where $L$ is the luminosity and $D$ is the characteristic dimension of the
emission region.
We adopt the following descriptions of $\lh$ (Z03),
\begin{equation}
\label{equ:lhz}
 \lhz(t) = A r(t)^{-2} \left[1-\sqrt{6 \over r(t)}\right],
 \ \ \ \ 0<t<\tmax
\end{equation}
which describes the dissipation of gravitational energy in an accreting ring.
Initial and final radial positions of the
propagating structures are denoted $\rstart$ and $\rend$, respectively.
$\tmax$ is the duration of a flare, computed assuming motion of the emission
region at a fraction of free-fall velocity (Z03).

The cooling of the plasma is described by cooling compactness, $\lsoft$.
In the basic model the main source of soft photons is reprocessing of hard
X--rays, which is described as (Z03),
\begin{equation}
 \label{equ:lsoft}
 \lsoft(t) = N C(t) \lh(t), 
\end{equation}
where $C(t)$ has the meaning of a covering factor of the reprocessing 
matter. It is described by
\begin{equation}
 \label{equ:refl}
   C[r(t)] = \left\{
 \begin{array}{cc}
             1                           & \mbox{for } r(t) \ge \rtr \\
  \left[{r(t)/\rtr}\right]^{\gamma}& \mbox{for } r(t) < \rtr,
 \end{array}
 \right.
\end{equation}
i.e.\ it equals 1 when the emitting structure is above the cold disc, 
but it gradually decreases to 0 approaching the black hole. 
The decline of $C(r)$  below the truncation radius, $\rtr$,
has indeed to be rather gradual, as discussed in some
detail in Z03; specifically the exponent is assumed $\gamma=2$.

As indicated above, the idea for generating a QPO is then to impose 
an additional 
periodic modulation of one or more of the above time dependencies. 
For example, if the QPO were a result of oscillations of the accretion disc,
then irrespectively of physical origin of the oscillations, they could
affect the flux of soft photons, or the temperature of the 
soft emission, $T_0$, 
or the amplitude of the reflected component, $R$, or some combinations of 
the above. This will produce a definite signature in, for example, energy
dependence of QPO strength, or, related quantity, energy spectra computed
at the QPO frequency.

The Comptonized spectra are computed with the {\sc eqpair} code 
(Coppi 1999). The code computes thermal and pair balance of a spherical plasma
cloud, given the heating rate of the plasma and cooling rate by soft photons,
and other necessary parameters. It then computes the spectrum of emerging radiation.
In our case, at each time step the heating and cooling compactnesses are calculated
from eq.~(\ref{equ:lhz}) and (\ref{equ:lsoft}) with the additional periodic
modulation as detailed below, and the spectrum of emergent radiation
is computed.
The reprocessed component is then added to it, with relative amplitude,
$\Or=C(t)$, where $\Or$ is the solid angle of the reprocessor from the X--ray 
source. Narrow Fe \Ka fluorescent
line is added to the Compton reflected continuum. Its energy is fixed at
6.4 keV, while its equivalent width (EW) depends of the  continuum slope 
(\.{Z}ycki \& Czerny 1994; Z03), roughly covering the range of 1 to 1.3 keV.

\begin{figure}
 \epsfxsize = 0.5\textwidth
 \epsfbox[18 260 600 700]{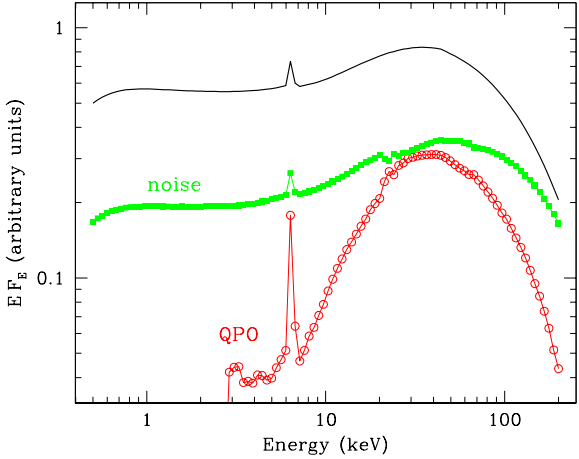}
 \caption{Energy spectra from a test case of modulation of reflection amplitude
only (Sec~\ref{sec:test}). Solid curve is the time averaged spectrum, squares
(green online) is the r.m.s.\ spectrum of the broad band noise component, while
circles (red online) is the r.m.s.\ spectrum of the QPO. The QPO was modeled 
by varying the amplitude of reflection only, and its r.m.s.\ spectrum reproduces
the shape of the reflected component.
\label{fig:test}}
\end{figure}

A sequence of spectra is computed with total duration of 256 sec.\ with time step 
of 1/32 sec. It is then analysed using standard tools of Fourier analysis.
The $f$-spectra are computed as described in 
Revnivtsev et al.\ (1999) and \.{Z}ycki (2002). An implicit assumption here is that
the  radius of the flaring region is constant in time, that is, the
flare luminosity is represented by the compactness parameter.
One may imagine more complex models where both $D$ and $L$ vary, which may lead
to luminosity--spectrum relations different than those presented in this paper.

\section{Results}
 \label{sec:results}

Computations are performed for a $10\,\MSun$ black hole system. 
Flare duration (travel time), $\tau$, is 
generated from a probability distribution $P(\tau) \propto \tau^{-1}$ between
$\taumin=0.005$ sec and $\taumax=2$ sec, so that the PDS roughly matches PDS
of Cyg X-1 (Z03). Initially, $\rstart$ is set to $100\,\Rg$ but it is set
smaller for short flares, so that the propagation velocity does not exceed the
free-fall velocity. The normalization constant in eq.~(\ref{equ:lhz}) is such that
the maximum of $\lh(t)=100$. The parameters are chosen so that the time average 
spectra are rather hard: normalization constant in 
eq.~\ref{equ:lsoft}, $N=0.4$, and $\rtr=20\,\Rg$, which give final spectrum with
$\Gamma \approx 1.8$. The QPO frequency is assumed 1 Hz.

\begin{figure*} 
 \begin{center}
 \parbox{\textwidth}{
  \epsfysize = 9.cm
  \epsfbox[18 390 600 700]{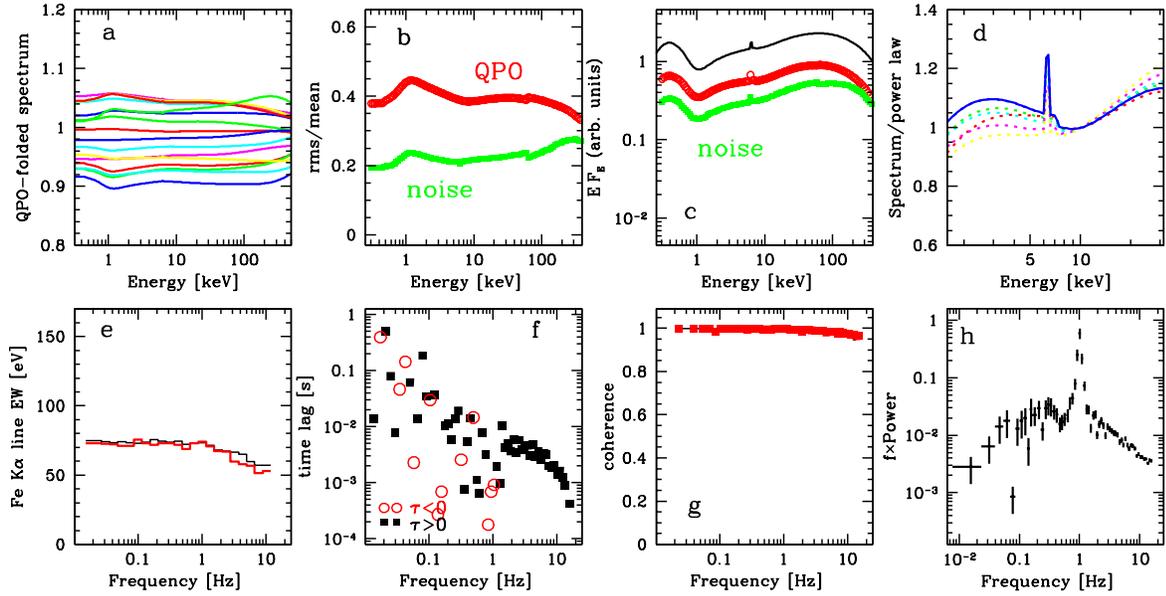}
}
 \caption{
 Simulation results for the case of quasi-periodic modulation of the heating rate,
with coherent response of the cooling rate (see Sec.~\ref{sec:lhresp}).
(a) shows the energy spectra folded with the period of the QPO, 
$\Tqpo=2\pi/\fqpo$, and divided by the time average spectrum.
The spectra show no clear spectral evolution during the QPO.
(b) shows the frequency-integrated rms
variability as a function
of energy for both the broad band noise component (excluding the QPO; solid squares, green
online) and the QPO (circles; red online). The QPO rms is basically independent of energy, 
as a consequence of the lack of spectral variability, while the noise rms
increases somewhat with energy.
(c) -- energy spectra: time averaged (solid curve), the variable component (filled
squares; green online).
(d) shows $f$-resolved spectra, computed in
the following ranges of $f$: 0.04--0.06 Hz, 0.4--0.7 Hz,
0.9--1.3 Hz (QPO; blue online), 1.7--2.5 Hz, 4.1-6.1 Hz and 11--16 Hz.
There is a weak overall
trend for hardening of the spectra with $f$ and the QPO spectrum 
is not much different than
for other values of frequency. 
(e) shows the EW of the Fe \Ka line as a function of $f$ (thick histogram; red
online), comparing it to the basic case of no QPO (thin histogram). 
Again, in the considered case there is no
additional signal at $\fqpo$. 
 (f) and (g) show results of 
cross-correlation analysis of light curves at 3 keV and 9 keV. 
The time lags show a minimum at $\fqpo$, which seems to be the only
signature of the QPO. The coherence function (g) is equal to 1 in almost entire 
range of $f$.
(h) shows the PDS from light curve at 9 keV.
\label{fig:reslh1}}
 \end{center}
\end{figure*}

\subsection{A test case -- modulation of reflection amplitude}
\label{sec:test}

Let us consider first a simple case of modulation of amplitude of the reprocessed
component, $\Or$. That is, we use un-modified eqs.~(\ref{equ:lhz}), (\ref{equ:lsoft}) and 
(\ref{equ:refl}) to describe heating and cooling of the plasma, but the reflection
amplitude is
\begin{equation}
 \Or(t) = \Aqpo \sin^2(2\pi\fqpo t + \phiqpo),
\end{equation}
with $\Aqpo=2$ and $\fqpo=0.5$ Hz.
We call it a test case because it is not a self-consistent situation 
in our model, since  $\Or$ should equal to the feedback function $C(t)$.
It does, nevertheless, demonstrate well the meaning of the r.m.s.\ spectrum. 
The r.m.s.\ spectra 
were obtained by first fitting to the PDS in each energy channel
a model consisting of a doubly broken power law,
\begin{equation}
 P(f) = N_P \times \left\{
\begin{array}{ll}
  \quad\ f^0, & \mbox{if  }  \quad\quad\quad f \le \fba \\
  N_1 f^{-1}, & \mbox{if  } \fba < f \le \fbb \\
  N_2 f^{-2}, & \mbox{if  } \quad\quad\quad f>\fbb
\end{array}
\right.
\end{equation}
(where $N_1=\fba$ and $N_2=\fba \fbb$ ensure continuity of $P(f)$),
with a Lorentzian QPO feature. Fitted parameters
are the overall normalization $N_P$, two break frequencies $\fba$, $\fbb$ and 
normalization and width of the Lorentzian. The rms$(E)$ are
then simply computed as integrals over $f$ of the continuum $P(f)$ and the 
QPO. The QPO r.m.s.\ spectrum plotted in Fig.~\ref{fig:test} reproduces the 
energy spectrum of the process which generated the QPO.

\subsection{Modulation of the heating rate}

Here we consider periodic modulation of the heating rate,
$\lh$. The motivation for this case comes from the description of broad band
X--ray power spectra as a sum of a number of Lorentzians 
(e.g., Miyamoto et al.\ 1991; Nowak 2000).
If taken seriously, it may be interpreted that  X--rays (not only QPO)
are emitted in the form of damped or forced oscillators. One might then 
imagine that a QPO is one of those oscillatory modes, of rather higher quality
factor than oscillations producing the broad band noise.
Furthermore, Maccarone, Coppi \& Poutanen (2000) 
demonstrated that in the low/hard state the flares have to be
driven by the heating rate rather than cooling flux of soft photons.
Specifically, we assume then
\begin{equation}
\label{equ:lheat}
 \lh(t-t_0) = \lhz(t-t_0)\left[1 + \Aqpo \sin( 2\pi \fqpo t + \phiqpo)\right],
\end{equation}
where $t_0$ is start time of a given flare. The time $t$ is a global time,
thus the above formula describes really a background modulation acting on a 
collection of  flares, whose start time and durations are randomly generated 
from
appropriate probability distributions (Z03). If the phase, $\phiqpo$, was
constant, the modulation would be a perfectly periodic, thus producing a narrow
feature in PDS. In order to produce a quasi-periodic
modulation, we change the phase abruptly at random moments in time. 
Such a procedure is motivated by results of Morgan, Remillard \& 
Greiner (1997), who found random walk of QPO phase in GRS~1915+105.
Parameters $\Aqpo$ and $\fqpo$ are constant.

Of crucial importance  is the response of the soft flux, $\lsoft$, 
to the modulation of $\lh$. In the basic model of Poutanen \& Fabian (1999) the
soft flux comes only from reprocessing of hard X--rays, which means that
the former must respond coherently to the variations of the latter. 
This assumption was made so that the model reproduces the spectral slope --
reflection amplitude correlation observed in low/hard state 
(Zdziarski, Lubi\'{n}ski \& Smith 1999). However, in soft states the disc
emission usually dominates the hard X-ray emission, therefore we have to allow
for the possibility that they are at least partially independent.
Consequently, we consider also the situation where the soft flux might 
respond only
to the flare of radiation, described by $\lhz(t)$, but it does not
respond to the quasi-periodic component.

\subsubsection{Cooling rate responding to QPO}
\label{sec:lhresp}

Here we assume that the flux of soft photons cooling the plasma is related
to the heating as
\begin{equation}
 \label{equ:lslh1}
 \lsoft(t) = N C(t) \times \lh(t),
\end{equation}
that is, the soft flux comes from reprocessing of the entire flux of
hard X--rays, including the quasi-periodic modulation. We assume that 
any delay in the response of $\lsoft(t)$ is negligible. This
is certainly correct for QPO of considered frequency, $f \le 10$ Hz, 
since the light travel time through $10\RS$ is only $10^{-3}$ s.

\begin{figure*} 
 \begin{center}
 \parbox{\textwidth}{
  \epsfysize = 9.0 cm
  \epsfbox[18 390 600 700]{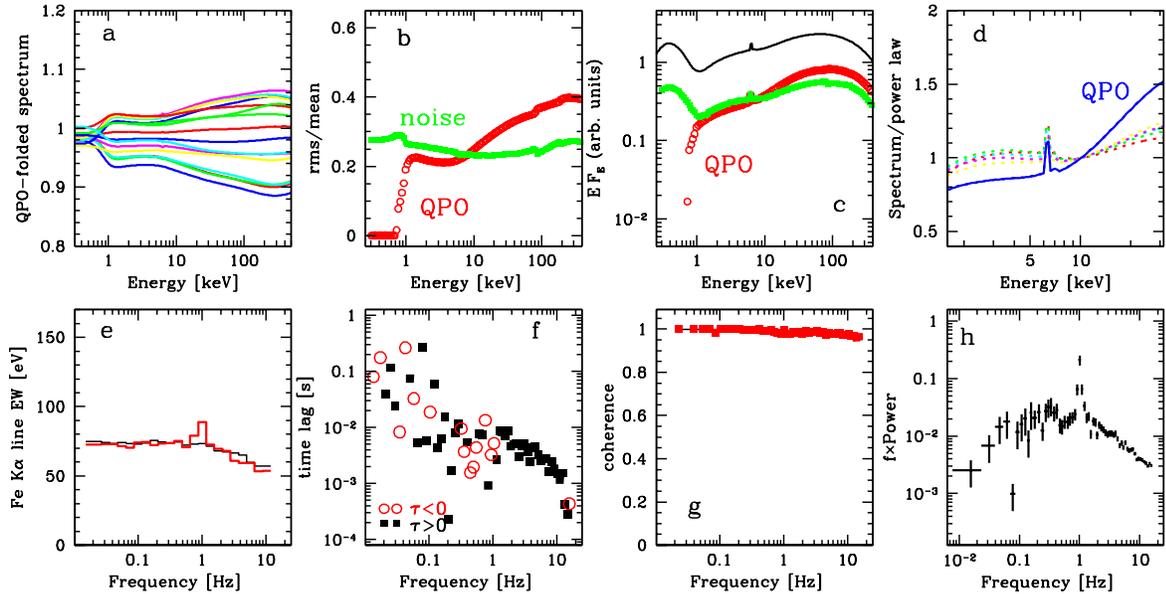}
}
 \caption{
 Results for the case of quasi-periodic modulation of the heating rate,
with no response of the cooling rate (see Sec.~\ref{sec:lhnoresp}). The QPO 
spectrum is harder here than the $f$-spectra at neighboring frequencies. 
It is also harder than the time averaged spectrum.
\label{fig:reslh2}}
 \end{center}
\end{figure*}

The results are presented in Fig.~\ref{fig:reslh1}. Panel (a) shows
the energy spectra folded with the period of the QPO, $\Tqpo=2\pi/\fqpo$,
and divided by the time average spectrum.
No clear spectral evolution during the QPO is seen. This is obviously
a consequence of eq.~\ref{equ:lslh1}, since the spectral slope depends
only on the ratio $\lh/\lsoft$, which is unaffected by the QPO.
Panel (b) shows the frequency-integrated r.m.s.\ variability as a function
of energy for both the broad band noise component (excluding the QPO)
and the QPO. This is computed as explained in Sec.~\ref{sec:test}.
The QPO rms$(E)$ is basically independent of energy, again
as a consequence of the lack of spectral variability. The noise rms$(E)$
increases somewhat with energy, because of the spectral evolution (hardening)
during each flare. 
Panel (c) shows energy spectra in $E\,F_E$ representation: time averaged spectrum,
r.m.s.\ spectrum of the variable component and the QPO r.m.s.\ spectrum. 
The two latter spectra
are simply the rms$(E)$ dependencies from previous panel multiplied by time averaged
count rate $F_E$.
Panel (d) show the $f$-spectra computed in the following ranges of $f$: 
0.04--0.06 Hz, 0.4--0.7 Hz, 0.9--1.3 Hz (QPO), 1.7--2.5 Hz, 4.1-6.1 Hz and 
11--16 Hz. There is a weak overall
trend for hardening of the spectra with $f$ and the QPO $f$-spectrum 
(i.e.\ the spectrum at $f$ around $\fqpo$) is not much different than
for other values of frequency\footnote{The QPO $f$-spectrum is constructed from
total $P(f)$ around $\fqpo$, i.e., containing contribution both from the broad band
noise and the QPO, while the previously considered QPO r.m.s.\ spectrum was constructed
from the component describing the QPO only}. 
Panel (e) shows the equivalent width (EW)
of the Fe \Ka line as a function of Fourier frequency, comparing it
to the basic case of no QPO. Again, in the considered case there is no
additional signal at $\fqpo$. Panels (f) and (g) show results of 
cross-correlation analysis of light curves at 3 keV and 9 keV. 
By construction, the model produces hard X--ray time lags. The time lags
show what seems to be the only signature of QPO in the considered case,
namely the lags show a clear minimum at $\fqpo$. The coherence function
is equal to 1 in almost entire range of $f$, in particular there is 
no signature of QPO here. Finally, (h) shows the PDS from light curve
at 9 keV.

\subsubsection{Cooling rate not responding to QPO}
\label{sec:lhnoresp}

In this case we assume that the soft flux comes from reprocessing of 
the non-periodic flare of hard X--rays only,
\begin{equation}
 \label{equ:lslh2}
 \lsoft(t) = N C(t) \times \lhz(t).
\end{equation}

It is not entirely clear how this situation may be realized geometrically,
since it would probably require a specifically anisotropic emission.
It is nevertheless an interesting one to examine because of clear 
observational signatures.
Results, plotted in Fig.~\ref{fig:reslh2} show an obvious pattern of 
spectral variability during the QPO: the  brighter the spectra the harder 
they are\footnote{Note that our model produces a broad band PDS with a QPO,
which means that individual spectra may not obey the trends; the trends
are visible as the shapes of envelopes of QPO period-folded spectra.}. 
A consequence of this is that the QPO rms increases with energy, and
the QPO energy spectrum is much harder than the other $f$-spectra.
The harder QPO spectrum produces somewhat larger EW of the \Ka line
at $\fqpo$. Similarly to the previous case, the 9 keV vs.\ 3 keV timelags 
show some sign changing around $\fqpo$, and the coherence function remains
equal to 1 at all $f$.

\begin{figure*} 
 \begin{center}
 \parbox{\textwidth}{
  \epsfysize = 9.0 cm
  \epsfbox[18 390 600 700]{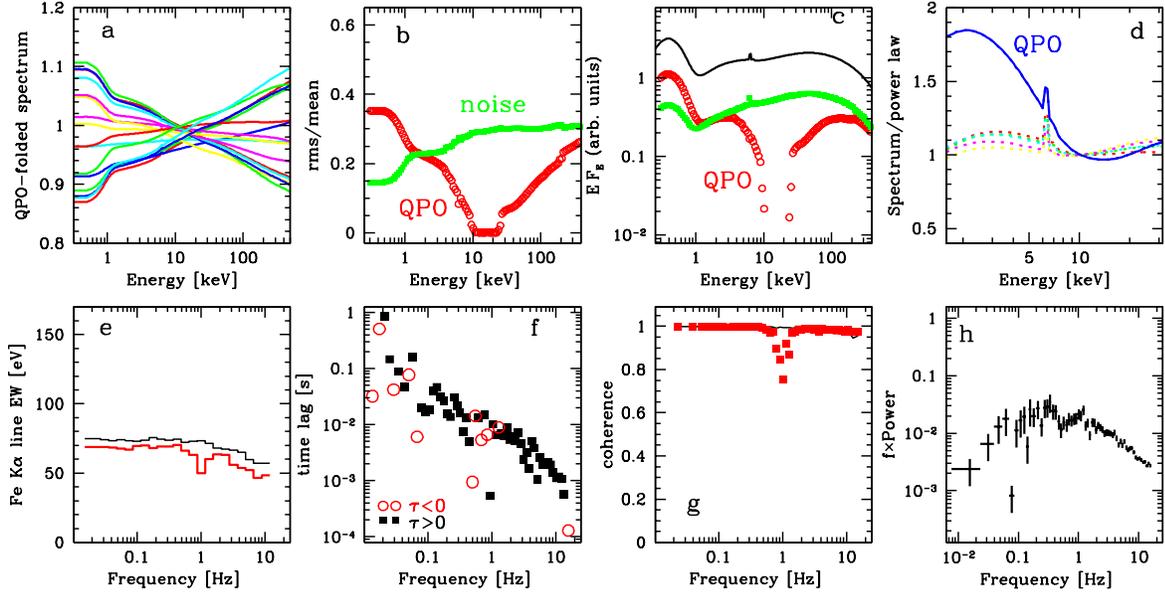}
}
 \caption{
 Results for the case of quasi-periodic modulation of the soft cooling flux
(see Sec.~\ref{sec:lstconst}) with constant temperature of the seed photons.
See caption to Fig.~\ref{fig:reslh1} for detailed description of panels and 
curves.
Characteristic pivoting behaviour resulting from modulations of $\lsoft$ gives
broad minimum in rms$(E)$ at $\sim 10$ keV and, consequently, softer QPO 
spectra and a minimum in \Ka line EW$(f)$ at $f=\fqpo$.
\label{fig:resls}}
 \end{center}
\end{figure*}

\subsection{Modulation of the soft flux}
\label{sec:lsoft}

Here we assume that the flux of soft photons is additionally modulated
in quasi-periodic manner. Again, we consider two situations: modulation of 
the flux with constant temperature of soft seed photons, and modulation of
the flux with corresponding change of the soft photons temperature.

\subsubsection{Constant soft photons temperature}
\label{sec:lstconst}

We assume that the temperature of the soft seed photons is constant in time,
while their flux is additionally modulated. 
Specifically, $\lh(t)=\lhz(t)$ 
(eq.~\ref{equ:lhz}) and
\begin{equation}
 \lsoft(t-t_0) = N C(t-t_0) \lh(t) +\Aqpo [\sin(2\pi\fqpo t+\phiqpo)]^2,
\end{equation}
where the square of the sine function assures that $\lsoft>0$.
Such form of $\lsoft$ could correspond to accretion disc pulsations
modulating the intrinsic (rather than reprocessed) flux of soft photons.
We now assume $\fqpo=0.5$ Hz, so that the QPO is still at 1 Hz in PDS.

Spectral variability driven by the soft flux produce a characteristic
pivoting behaviour, with the pivot energy 
intermediate between $T_0$ and $\kT$.
This is evident in our simulations presented in Fig.~\ref{fig:resls},
where the pivot energy is about 20 keV. As a consequence, the rms$(E)$ has
a broad minimum around 20 keV. The QPO spectrum is in consequence much
softer than the other $f$-spectra, at least up to $\approx 10$ keV. This
in turn explains the small EW of \Ka line at the QPO frequency.
Time lags between 3 keV and 9 keV light curves do not show any clear
feature at $\fqpo$, but a sharp drop of the coherence function at
$\fqpo$ is observed. This is caused by the existence of two independent 
Fourier components at $\fqpo$: one from the broad band noise, the other from 
the QPO. The lack of QPO in PDS from the 9 keV light curve is consistent with 
the presence of minimum in rms$(E)$.

\subsubsection{Modulation of the seed photons temperature}
\label{sec:lstvar}

We assume that the temperature of the soft photons, $\Tsoft$, is modulated as 
\begin{equation}
  \Tsoft(t) = T_0 [1 +  \Aqpo \sin( 2\pi \fqpo t + \phiqpo)],
\end{equation}
and the corresponding modulation of the soft flux is given by
\begin{equation}
 \lsoft(t) = N C(t) \lh(t) + N \left({\Tsoft \over T_0}\right)^4,
\end{equation}
where we assume $\Aqpo=0.1$ and $\fqpo=1$ Hz. The above form 
describes two sources of soft photons: in addition to the reprocessed photons,
there is an additional flux from  periodically modulated intrinsic dissipation. 
We assume that the geometrical dependence of the modulated flux is different
than that of the  reprocessed flux and that the covering factor $C(t)$ does
not appear in the second term in the expression for cooling compactness.

\begin{figure*} 
 \begin{center}
 \parbox{\textwidth}{
  \epsfysize = 9.0 cm
  \epsfbox[18 390 600 700]{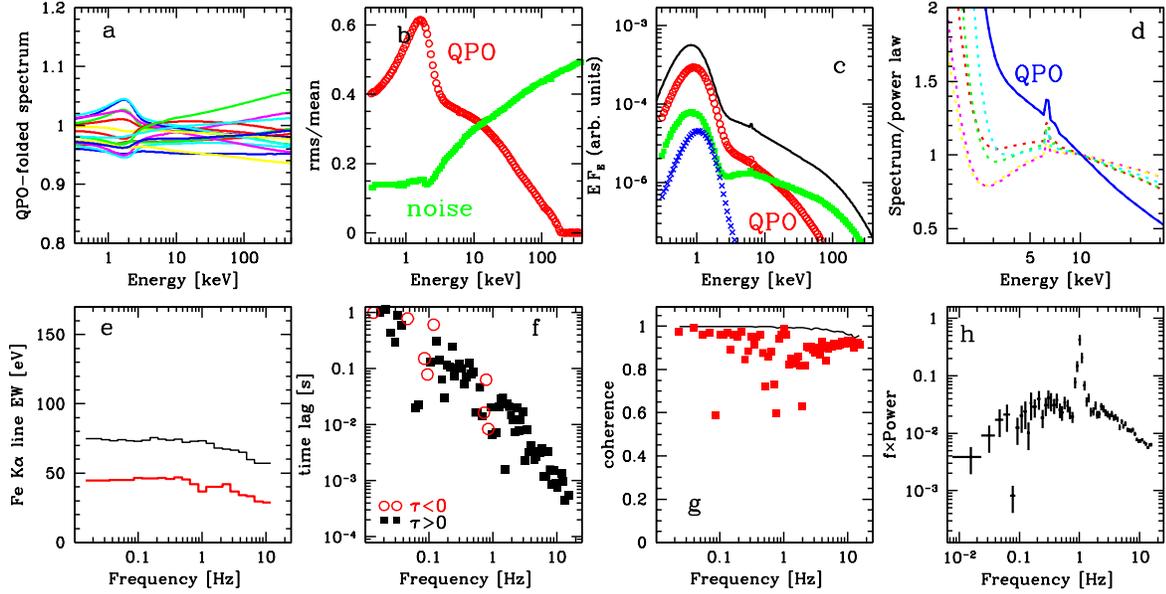}
}
 \caption{
 Results for the case of quasi-periodic modulation of the soft cooling flux,
with corresponding change of seed photons temperature (Sec.~\ref{sec:lstvar}),
are quite different than those with constant $\Tsoft$ (Fig.~\ref{fig:resls}).
Maximum variability is now at low energies, and the QPO spectrum is very soft.
Importantly, this modulation produces also a harmonic of the QPO, although
only at low energies (it is no longer visible at 9 keV). Crosses in panel (c) 
(blue online) show the spectrum of the QPO harmonic. Coherence function (g) shows
very complex behaviour: it seems to have a maxima at $\fqpo$ and $2\fqpo$, 
with values much lower than 1 around these two frequencies.
\label{fig:reslst}}
 \end{center}
\end{figure*}

The results (Fig~\ref{fig:reslst}) are rather different than those for previous 
case of constant $\Tsoft$. Strongest variability is now at low energies, around
$E=k \Tsoft$, and its amplitude decreases with energy. Consequently, the QPO
spectra are soft and the Fe \Ka line EW is lower than our reference case.
It does not show any clear feature at $\fqpo$. Time lags do show some sign
changes around $\fqpo$, while the coherence shows a complex behaviour,
with local minima around $\fqpo$, but a maximum equal to 1 at $\fqpo$. There
is also a harmonic QPO at $2\fqpo$, which is related to the fact that the
flux modulation is not a simple sinusoidal modulation. The harmonic is strong only
at the energies where the soft flux appears.

\subsection{Modulation of the covering factor of the reprocessor}
\label{sec:refl}

\begin{figure*} 
 \begin{center}
 \parbox{\textwidth}{
  \epsfysize = 9.0 cm
  \epsfbox[18 390 600 700]{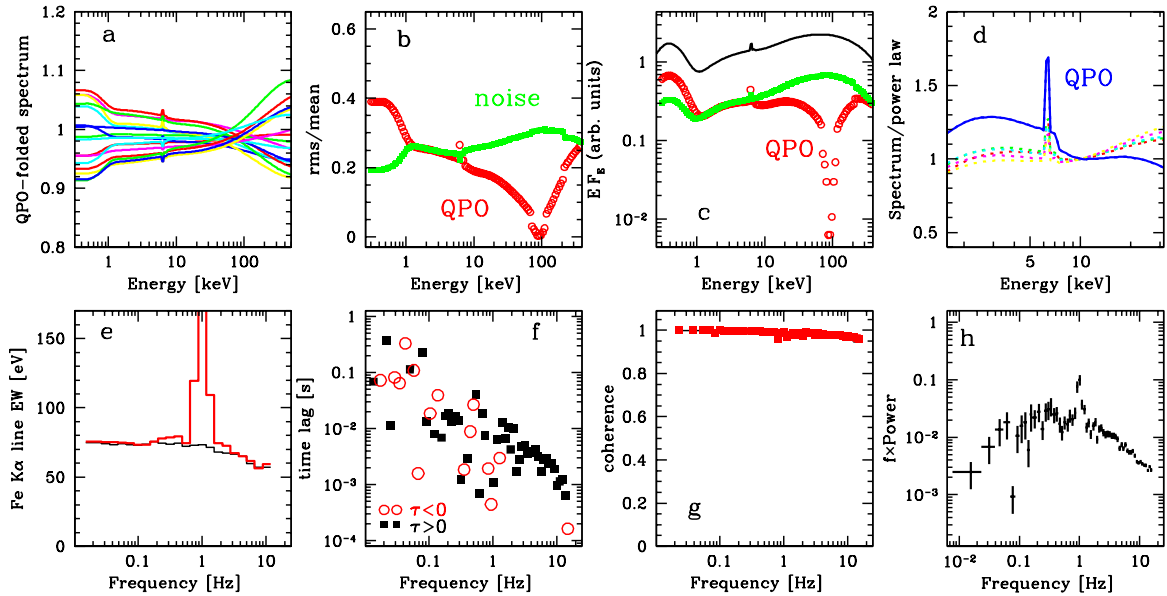}
}
 \caption{
 Results for the case of quasi-periodic modulation of the covering factor of the
cold reprocessing matter, $C(t)$, (see Sec.~\ref{sec:refl}), which controls both 
the feedback 
between heating and cooling of the hot plasma (eq.~\ref{equ:lsoft}), and the
amplitude of reflection, $\Or$. Strong Fe \Ka line present in the QPO spectra is
the most characteristic feature of this model.
\label{fig:resrefl}}
 \end{center}
\end{figure*}

Here we assume that the covering factor of cold matter is additionally
quasi-periodically modulated,
\begin{equation}
 C(t) = C_0(t) [1 + \Aqpo\sin(2\pi\fqpo t + \phiqpo)],
\end{equation}
with $\Aqpo=0.5$, $\fqpo=1$ Hz, and $\phiqpo$ generated as described
earlier.
Both the heating rate and cooling rate are described by their
basic equations, eq.~\ref{equ:lhz} and eq.~\ref{equ:lsoft}, respectively.
Compared to the previous case of modulation of $\lsoft$, now the amplitude 
of the reprocessed component (Compton reflected continuum
with consistent Fe \Ka line) varies as well, $\Or(t)=C(t)$.
This scenario may correspond to disc warping due to, e.g., Lens--Thirring
precession (e.g., Miller \& Homan 2005). 
Results of computations are presented in Fig.~\ref{fig:resrefl}. 
The QPO-folded energy spectra show clear pivoting around 30--50 keV.
The variability of the \Ka line with the QPO phase  is noticeable even in 
those folded spectra. The QPO rms$(E)$ dependence shows an increase of
variability rms at the line energy, and a broad minimum around 20--50 keV.
The QPO spectrum is then softer than the other $f$-spectra, while
the EW of the \Ka line has a sharp maximum at $\fqpo$. This is a clear
difference with the previous case of $\lsoft$ modulation, where a soft
QPO spectrum was accompanied by a weak line. Here, 
the correlation between the spectral slope and amplitude of reflection
is obeyed in the $f$-spectra, because of eq.~\ref{equ:lsoft}. Thus
a soft continuum is accompanied by a strong reflection component, hence
strong \Ka line. The 3 vs.\ 9 keV time lags appear to reverse sign at 
$\fqpo$ (but are positive on both sides of it), but there is only very weak
loss of coherence between the two light curves at $\fqpo$.

\subsection{Modulation of absorption column}

One might imagine that modulation of the observed flux could be produced
simply  as a result of obscuration by an absorber. Here the results fall
into either of two extreme categories. If the variable obscuration affects
the source spectrum, the results are very complex. For example, 
the power spectra contain
a number of strong harmonics at $2\times$, $3\times$, \dots of the fundamental
frequency. Also the coherence function then shows a number of minima at the
same $f$'s. This is obviously because the way the spectra are distorted by 
absorption is not linear in $\NH$, thus the complexity appears even if $\NH$
is modulated periodically. On the other hand, if the absorber is very optically
thick $\tau_{\rm T} \gg 5$, so that practically no photons are transmitted
and the only variable factor is the covering factor, then only the luminosity
is modulated, but no spectral signatures are observed.

\section{Discussion}
\label{sec:discuss}

We have constructed models of X--ray QPO in accreting compact systems, 
and computed specific energy-spectral properties of the QPO in each model.
We have assumed general radial propagation model, where the X--ray emission
is produced by structures moving radially towards the central black hole
(Kotov et al.\ 2001; Z03).
Assuming that the relevant radiative process is thermal Comptonization,
we considered cases when the quasi-periodically modulated factor is 
plasma heating rate, 
cooling rate by soft photons, amplitude of the reprocessed component 
(feedback factor between the heating rate and cooling rate), or the 
temperature of soft photons. For each case we have constructed a set of
observable quantities, which may be compared to good quality data.
These are generally of two types: energy dependencies at various Fourier 
frequencies,
and Fourier frequency dependencies, showing possible signatures at $\fqpo$.
QPO amplitude as a function of $E$ is the obvious characteristic, but it is
really a powerful tool for constraining models only when used in conjunction
with similar dependencies at other $f$. Such Fourier spectra 
show distinct signatures at $\fqpo$ in most considered 
models. The QPO spectrum is harder than other $f$-spectra when $\lh$ is 
modulated and $\lsoft$ does {\em not\/} respond fully to that modulation.
The QPO spectrum is softer than other $f$-spectra if the modulated factor
is $\lsoft$ or the reflection amplitude, $\Or$ (assuming the latter controls
the feedback between $\lh$ and $\lsoft$). The difference between modulation
of $\lsoft$ and $\Or$ is in the behaviour of Fe \Ka line as a function
of $f$: for $\lsoft$ modulation it shows local minimum at $\fqpo$, while it shows
local maximum at $\fqpo$ for $\Or$ modulation. The coherence function also
differs between the case of $\lsoft$ and $\Or$ modulation: it shows a dip
at $\fqpo$ (weaker coherence) in the former case only.

The significance of the presented models is in offering possibilities
of providing hints as to the physical models of QPO. The basic question is whether
the modulation occurs as a result of instabilities in the hot plasma 
or in the cold disc. The former would produce modulation of the heating rate, while
the latter modulation of the cooling rate, soft photons temperature, reflection 
amplitude, or some combination of these. Candidates for the cold disc instabilities
have been discussed in literature for some time now (review in Kato, Fukue
\& Mineshige 1998), but no work was done on consequences of these on 
X--ray spectra. This is of course not surprising because physical models of 
transitions between and interactions of the cold disc and hot corona are at 
a very early stages of development (R\'{o}\.{z}a\'{n}ska \& Czerny 2000, 
Giannios \& Spruit 2004 and references therein).

Models presented in this paper certainly do not cover all possibilities.
One can imagine other, more complex scenarios. In particular, 
we have assumed here purely thermal Comptonization, while most spectra of sources
in bright soft states require hybrid thermal/non-thermal plasmas (Gierli\'{n}ski 
\& Done 2003 and references therein). Fits to such spectra
usually imply that both direct heating and injection of energetic non-thermal
particles contribute to the energy input. One can then imagine that a modulation
might affect one of those channels, or their relative importance. We discuss
some of such models in Sobolewska \& \.{Z}ycki (2005) as a possible
description of low frequency QPO in XTE~J1550-564 in anomalous very high state.

Generally, observational data on QPO energy dependencies from a number of black 
hole binaries  point to a complex picture (Sobolewska \& \.{Z}ycki 2005),
although certain correlations do appear.  The QPO r.m.s.\ spectra of sources
in hard states seem to be softer than the time averaged spectra, while the
QPO spectra of sources in soft state appear to be harder than time averaged
spectra. According to our analysis, this would point to driving the modulation
by plasma heating in soft state, while in the hard state the driver would be
related to changes in the cold disc.
This would correspond to the modes of long-term variability
discussed for Cyg X-1 (Zdziarski et al.\ 2002), perhaps suggesting that the
low frequency QPO (which appear preferentially during state transitions;
Rutledge et al.\ 1999), are simply faster and more coherent version of the
general variability pattern operating in a given spectral state. We discuss
this in more detail in Sobolewska \.{Z}ycki (2005).

\section{Conclusions}

Signatures of quasi-periodic modulation of physical parameters in spectral
and timing observables are complex, depending on details of a particular model.
The QPO r.m.s.\ spectra can be harder than time average spectra if the 
the plasma heating rate is modulated. The QPO r.m.s.\ spectra can be softer
than the time average spectrum if parameters of the cool disc are
modulated: flux of cooling soft photons or covering factor of the cold plasma.

\section*{Acknowledgments} 
 
This work  was partly supported by grants no.\  2P03D01225
and PBZ-KBN-054/P03/2001
from the Polish State Committee for Scientific Research (KBN).

{}



\begin{thebibliography}{}
 
 \bibitem[]{}
  Belloni T., Psaltis D., van der Klis M., 2002, ApJ, 572, 392
 \bibitem[]{}
   Coppi P. S., 1999, in Poutanen J., Svensson R., eds, ASP Conf Ser. Vol. 161.
   Astron. Soc. Pac., San Francisco, p.~375 (astro-ph/9903158)
\bibitem[]{}
  Cui W., Zhang S. N., Chen W., Morgan E. H., 1999, ApJ, L32
 \bibitem[]{}
   Done C., 2002, Philosophical Transactions of the Royal Society, 360, 1967
    (astro-ph/0203246)
 \bibitem[]{}
   Done C., Gierli\'{n}ski M., 2004, Prog. Theor. Phys. Supp., 155, 9 (astro-ph/0403546)
 \bibitem[]{}
   Done C., Wardzi\'{n}ski G., Gierli\'{n}ski M., 2004, MNRAS, 349, 393
\bibitem[]{}
  Giannios D. \& Spruit H. C, 2004, A\&A, 427, 251
\bibitem[]{}
  Gierli\'{n}ski M. \& Done C., 2003, MNRAS, 342, 1083
\bibitem[]{}
  Kaaret P., Piraino S., Bloser P. F., Ford E. C., Grindlay J. E., Santangelo A., 
    Smale A. P., Zhang W., 1999, ApJ, 520, L37
\bibitem[]{}
  Kalemci E., Tomsick J. A., Rothschild R. E., Pottschmidt K., Kaaret P., 2004, ApJ,
  603, 231
\bibitem[]{}
  Kato S., Fukue J., Mineshige S., 1998, Black-Hole Accretion Disks, Kyoto 
         University Press, Kyoto
 \bibitem[]{}
   Kotov O., Churazov E., Gilfanov M., 2001, MNRAS, 327, 799
\bibitem[]{}
  Kubota A., Done C., 2004, MNRAS, 353, 980
\bibitem[]{}
  Lee W. H., Abramowicz M. A., Klu\'{z}niak W., 2004, ApJ, 603, L93
 \bibitem[]{}
   Maccarone T. J., Coppi P. S.,  2003, MNRAS, 338, 189
 \bibitem[]{}
   Maccarone T. J., Coppi P. S., Poutanen J., 2000, ApJ, 537, L107
 \bibitem []{}
  Markowitz A., et al., 2003, ApJ, 593, 96
 \bibitem[]{}
    Miller J. M., Homan J., 2005, ApJ, 618, L107
 \bibitem[]{}
   Miyamoto S., Kimura K., Kitamoto S., Dotani T., Ebisawa K., 1991, ApJ, 383, 784
\bibitem[]{}
   Morgan E. H., Remillard R. A., Greiner J., 1997, ApJ, 482, 993
 \bibitem[]{}
   Nowak M. A., 2000, MNRAS, 318, 361
 \bibitem[]{}
   Poutanen J., 2001, AdSpR, 28, 267 (astro-ph/0102325)
 \bibitem[]{}
   Poutanen J., Fabian A. C., 1999, MNRAS, 306, L31 
\bibitem[]{}
   Psaltis D., 2001,  AdSpR, 28, 481 (astro-ph/0012251)
 \bibitem[]{}
   Revnivtsev M., Gilfanov M., Churazov E., 1999, A\&A, 347, L23
\bibitem[]{}
  Rodriguez J., Corbel S., Hannikainen D. C., Belloni T., Paizis A., Vilhu O.,
     2004, ApJ, 615, 416
 \bibitem[]{}
   R\'{o}\.{z}a\'{n}ska A., Czerny B., 2000, A\&A, 360, 1170
\bibitem[]{}
   Rutledge R. E. et al., 1999, ApJS, 124, 265
\bibitem[]{}
   Sobolewska M., \.{Z}ycki P. T., 2004, A\&A, 400, 553
\bibitem[]{}
   Sobolewska M., \.{Z}ycki P. T., 2005, in preparation
 \bibitem[]{}
   Stern B. E., Svensson R., 1996, ApJ, 469, L109
 \bibitem[]{}
   Uttley P., McHardy I. M., 2001, MNRAS, 323, L26
 \bibitem[]{}
  van der Klis M., 1995, in Lewin W. H. G., van Paradijs J., 
      van den Heuvel E. P. J., eds, X--ray binaries, Springer,
       Berlin, p.\ 157
 \bibitem[]{}
  van der Klis M., 2001, ApJ, 561, 943
\bibitem[]{}
 Wijnands R., 2001,  AdSpR, 28, 469 (astro-ph/0008096)
 \bibitem[]{}
   Zdziarski A. A., Gierli\'{n}ski M., 2004, Prog. Theor. Phys. Supp., 155, 99
 \bibitem[]{}
   Zdziarski A. A., Lubi\'{n}ski P., Smith D. A., 1999, MNRAS, 303, L11
 \bibitem[]{}
   Zdziarski A. A., Poutanen J., Paciesas W. S., Wen L., 2003, ApJ, 578, 357
 \bibitem[]{}
    \.{Z}ycki P. T., 2002, MNRAS, 333, 800
 \bibitem[]{}
    \.{Z}ycki P. T., 2003, MNRAS, 340, 639 (Z03)
 \bibitem[]{}
    \.{Z}ycki P. T., 2004, MNRAS, 351, 1180
 \bibitem[]{}
    \.{Z}ycki P. T., Czerny B., 1994, MNRAS, 266, 653

\label{lastpage}

\end{thebibliography}
\end{document}